\documentclass[a4paper,fleqn,usenatbib]{mnras}
\usepackage{float}
\usepackage{amsmath}
\usepackage{booktabs}
\usepackage{pdflscape}
\usepackage{txfonts}
\usepackage[T1]{fontenc}
\usepackage{ae,aecompl}
\usepackage{graphicx}
\usepackage{color,soul}

\title[Comparison of active and inactive galaxies in ZFOURGE]{ZFOURGE catalogue of AGN candidates: \\an enhancement of 160$\mu$m-derived star-formation rates in active galaxies to $z=3.2$ \thanks{This paper includes data gathered with the 6.5 meter Magellan Telescopes located at Las Campanas Observatory, Chile.}}

\author[M. J. Cowley et al.]{Michael J. Cowley,$^{1,2}$\thanks{E-mail: michael.cowley@students.mq.edu.au}
Lee R. Spitler,$^{1,2}$
Kim-Vy H. Tran,$^{3}$
Glen A. Rees,$^{1,4}$
\newauthor 
Ivo Labb\'e,$^{5}$
Rebecca J. Allen,$^{2,6}$
Gabriel B. Brammer,$^{7}$
Karl Glazebrook,$^{6}$
\newauthor 
Andrew M. Hopkins,$^{2}$
St\'ephanie Juneau,$^{8}$
Glenn G. Kacprzak,$^{6}$
\newauthor
James R. Mullaney,$^{9}$
Themiya Nanayakkara,$^{6}$
Casey Papovich,$^{6}$
Ryan F. Quadri,$^{5}$
\newauthor
Caroline M. S. Straatman,$^{5}$
Adam R. Tomczak,$^{3,10}$
and Pieter G. van Dokkum$^{11}$
\\
\\
$^{1}$Department of Physics and Astronomy, Macquarie University, NSW 2109, Australia\\
$^{2}$Australian Astronomical Observatory, PO Box 915, North Ryde, NSW 1670, Australia\\
$^{3}$George P. and Cynthia W. Mitchell Institute for Fundamental Physics and Astronomy, Department of Physics and Astronomy, Texas \\
$^{4}$CSIRO Australia Telescope National Facility, PO Box 76, Epping, NSW 1710, Australia\\
A\&M University, College Station, TX 77843, USA\\
$^{5}$Leiden Observatory, Leiden University, PO Box 9513, 2300 RA Leiden, The Netherlands\\
$^{6}$Centre for Astrophysics and Supercomputing, Swinburne University, Hawthorn, VIC 3122, Australia\\
$^{7}$Space Telescope Science Institute, 3700 San Martin Drive, Baltimore, MD 21218, USA\\
$^{8}$CEA-Saclay, DSM/IRFU/SAp, F-91191 Gif-sur-Yvette, France\\
$^{9}$Department of Physics and Astronomy, The University of Sheffield, Hounsfield Road, Sheffield, S3 7RH, UK\\
$^{10}$Department of Physics, University of California Davis, One Shields Avenue, Davis, CA 95616, USA\\
$^{11}$Department of Astronomy, Yale University, New Haven, CT 06520, USA
}

\date{Accepted XXX. Received YYY; in original form ZZZ}

\pubyear{2015}

\begin{document}
\label{firstpage}
\pagerange{\pageref{firstpage}--\pageref{lastpage}}
\maketitle

\begin{abstract}
We investigate active galactic nuclei (AGN) candidates within the FourStar Galaxy Evolution Survey (ZFOURGE) to determine the impact they have on star-formation in their host galaxies. We first identify a population of radio, X-ray, and infrared-selected AGN by cross-matching the deep $K_{\mathrm{s}}$-band imaging of ZFOURGE with overlapping multi-wavelength data. From this, we construct a mass-complete (log($M_{*}/M_{\odot}$) $\geq 9.75$), AGN luminosity limited sample of 235 AGN hosts over $z= 0.2-3.2$. We compare the rest-frame $U-V$ versus $V-J$ ($UVJ$) colours and specific star-formation rates (sSFRs) of the AGN hosts to a mass-matched control sample of inactive (non-AGN) galaxies. $UVJ$ diagnostics reveal AGN tend to be hosted in a lower fraction of quiescent galaxies and a higher fraction of dusty galaxies than the control sample. Using 160$\mu$m \textit{Herschel} PACS data, we find the mean specific star-formation rate of AGN hosts to be elevated by $0.34\pm0.07$ dex with respect to the control sample across all redshifts. This offset is primarily driven by infrared-selected AGN, where the mean sSFR is found to be elevated by as much as a factor of $\sim$5. The remaining population, comprised predominantly of X-ray AGN hosts, is found mostly consistent with inactive galaxies, exhibiting only a marginal elevation. We discuss scenarios that may explain these findings and postulate that AGN are less likely to be a dominant mechanism for moderating galaxy growth via quenching than has previously been suggested.
\end{abstract}


\clearpage

\section{Introduction}

There is mounting evidence demonstrating that supermassive black holes (SMBHs) play a fundamental role in the formation and evolution of galaxies over cosmic time. Previous work has found the mass of a SMBH is tightly correlated with various properties of its host's hot spheroidal bulge, including its luminosity \citep[e.g.][]{Kormendy:1995hc, Graham:2007ii, Sani:2011fr} mass \citep[e.g.][]{Magorrian:1998cs, Marconi:2003vg, Beifiori:2011di} and velocity dispersion \citep[e.g.][]{Gebhardt:2000ih, Gultekin:2009hj, Graham:2011hc}. During periods of rapid accretion, the galactic nuclei of these systems can also release an immense amount of energy into the surrounding environment of the host galaxy \citep[e.g.][]{Kormendy:1995hc, Magorrian:1998cs}. As a result, theoretical simulations commonly invoke feedback from these active galactic nuclei (AGN) outflows to regulate the star-formation activity of galaxies \citep[e.g.][]{Ciotti:1997kd, Silk:1998up, Croton:2006ew}. The inclusion of a negative feedback mechanism helps resolve the overproduction of massive galaxies in simulations by heating or driving out gas to suppress star-formation. While observational evidence supports negative feedback via AGN-driven outflows \citep[e.g.][]{Nesvadba:2006jt, Feruglio:2010cy, Fischer:2010kz}, recent studies also point to the possibility of AGN producing positive feedback, whereby AGN outflows trigger star-formation by compressing cold dense gas. \citep[e.g.][]{Silk:2009fg, Elbaz:2009jb, Zinn:2013hb}. 

In order to reconcile these contradictory outcomes, the complex interplay between AGN activity and star-formation must be examined. Early studies, which tried to achieve this, relied on optical spectra to select AGN from large parent samples of galaxies. The main drawback of this approach was the restriction of low redshifts \citep[$z<0.3$;][]{Ho:2005en, Kim:2006kw, Salim:2007hq}. With cosmic AGN activity peaking at a similar epoch to cosmic star-formation ($z\sim 2$), these studies potentially miss a key phase of AGN evolution. 

More recent studies have pushed to higher redshifts by taking advantage of X-ray emission, which is an effective probe of AGN activity. Upon comparing X-ray AGN hosts to mass-matched reference galaxies, these studies yield results suggesting only minor or no difference in star-formation activity between the two samples \citep{Xue:2010dk, Santini:2012jx, Mullaney:2011ce, Rosario:2014jz}. However, by relying on X-ray selected AGN, these studies may also miss a key phase when AGN are hosted in dust-rich, X-ray obscured galaxies \citep{Sanders:1988fc}.

In this paper, we expand on this work by investigating the empirical connection between AGN activity and star-formation by selecting and analysing a diverse sample of AGN across a broad range of obscuration levels over $z= 0.2-3.2$. Our parent sample is the deep $K_{\mathrm{s}}$-band imaging of ZFOURGE ({\color{blue}Straatman et al.} {\color{blue}2015, submitted}), which not only grants us access to all galaxies types, but also allows us to probe to lower stellar masses and higher redshifts. 

To identify AGN, we cross-match the $K_{\mathrm{s}}$-band imaging with radio, X-ray and infrared (IR) datasets to allow the use of standard AGN selection techniques, and make use of rest-frame $UVJ$ colours to distinguish quiescent galaxies from star-forming galaxies. To gauge star-formation activity, we employ deep far-infrared (FIR) data (160$\mu$m) from the \textit{Herschel Space Observatory}. Our principal aim is to compare AGN hosts with a mass-matched sample of inactive galaxies, before discussing the implications of our results for understanding the connection between star-formation and AGN activity, as well as the impact AGN has on galaxy evolution.

This paper is structured as follows. In Sections 2 and 3, we describe the ZFOURGE and multi-wavelength data sets and AGN sample construction, while in Section 4 we outline our methodology to construct a mass-matched sample of inactive galaxies. In Section 5, we present our comparative analysis, before discussing the results and their implications in Section 6. Finally, we summarise our findings in Section 7. 

Throughout this paper, we use an AB magnitude system, a \citet{Chabrier:2003ki} IMF, and assume a $\Lambda$CDM cosmology with $\mathrm{H_0}$ = 70 km s$^{-1}$ Mpc$^{-1}$, $\Omega_M$ = 0.3, $\Omega_{\Lambda}$ = 0.7.

\section{ZFOURGE AND ANCILLARY DATA SETS}

\subsection{Galaxy Catalogues}

Our parent sample is comprised of galaxies identified in the ZFOURGE\footnote{\url{http://zfourge.tamu.edu}} survey, which covers three 11' x 11' pointings in the CDFS \citep{Giacconi:2002ef}, COSMOS \citep{Scoville:2007dl} and UDS \citep{Lawrence:2007hu} legacy fields. ZFOURGE uniquely employs deep near-IR imaging taken with five medium-band filters on the FourStar imager \citep{Persson:2013eo} mounted on the 6.5m Magellan Baade telescope. The imaging reaches 5$\sigma$ point-source limiting depths of $\sim$26 AB mag in $J_1, J_2, J_3$ and $\sim$25 AB mag in $H_{\mathrm{s}}$, $H_{\mathrm{l}}$, $K_{\mathrm{s}}$ \citep{Spitler:2012fk}. For galaxies at redshifts $z=1.5-4$, these filters bracket the rest-frame 4000\AA/Balmer breaks, resulting in well-constrained photometric redshifts within $\sigma (z)/(1+z) \approx 1-2\%$ \citep[e.g.][]{Kawinwanichakij:2014bn}. ZFOURGE is supplemented with existing data from CANDELS HST/WFC3/F160W \citep{Grogin:2011hx, Koekemoer:2011br, Skelton:2014do} and {\it Spitzer}/IRAC, as well as other ground-based imaging, to generate multi-wavelength catalogues spanning $0.3-8$ $\mu$m. Fluxes at wavelengths of the Infrared Array Camera \citep[IRAC;][]{Fazio:2004eb}, 3.6, 4.5, 6.8, and 8.0 $\mu$m are measured using the deblending approach described in \citet{Labbe:2006fo}. For further details on the acquisition, data reduction, and bands used to construct the ZFOURGE catalogues, see \citet{Tomczak:2014hw} and {\color{blue}Straatman et al.} ({\color{blue}2015, submitted}).

\subsection{Radio Data}

Following \citet{Rees:2015arXiv}, we cross-match ZFOURGE with published radio sources based on overlapping data from the Very Large Array (VLA). We use the VLA 1.4 GHz Survey of the Extended {\it Chandra} Deep Field South: Second Data Release of \citet{Miller:2013hp} for the ZFOURGE-CDFS field, the VLA-COSMOS Survey IV Deep Data and Joint catalogue of \citet{Schinnerer:2010je} for the ZFOURGE-COSMOS field, and the Subaru/XMM-Newton Deep Field-I 100 $\mu$Jy catalogue of \citet{Simpson:2006jh} for the ZFOURGE-UDS field. The minimum root-mean-square (RMS) sensitivity for each survey is 6, 10 and 100 $\mu$Jy/beam, respectively. Upon correcting for systematic astrometric offsets in each field, radio sources are cross-matched within a radius of 1$''$ of their $K_{\mathrm{s}}$-band counterparts. Of the 286 radio sources that overlap with the ZFOURGE fields, 264 were cross-matched with a $K_{\mathrm{s}}$-band counterpart. We visually inspect the remaining 22 sources and find 2 in the ZFOURGE-COSMOS field were missed due to confusion from complex extended structures (i.e. radio jets), with their recorded position offset from the galaxy core. The remaining 20 sources are considered candidate IR faint radio sources \citep[IRFS;][]{Norris:2006cy}, with a visual inspection yielding no identifiable counterparts in the $K_{\mathrm{s}}$-band images. Considering this, a total of 266 radio counterparts are found in the ZFOURGE $K_{\mathrm{s}}$-band images ($\sim$92$\%$ of all overlapping radio sources), with 119 in CDFS, 116 in COSMOS, and 31 in UDS. 

\subsection{X-Ray Data}

We cross-match ZFOURGE with published X-ray sources based on overlapping data from the {\it Chandra} and {\it XMM-Newton} space observatories. We use the {\it Chandra} Deep Field-South Survey: 4 Ms Source catalogue of \citet{Xue:2011bt} for the ZFOURGE-CDFS field (X11 henceforth), the {\it Chandra} COSMOS Survey I. Overview and Point Source catalogue of \citet{Elvis:2009io} for the ZFOURGE-COSMOS field (E09 henceforth), and the {\it Subaru/XMM-Newton} Deep Survey III. X-Ray Data of \citet{Ueda:2008dr} for the ZFOURGE-UDS field (U08 henceforth). The on-axis limiting flux in the soft and hard bands for each survey is $9.1\times { 10 }^{ -18 }\ \mathrm{and}\ 5.5\times { 10 }^{ -17 }\ \mathrm{erg\ {cm}^{-2}\ {s}^{-1}}$, $1.9\times { 10 }^{ -16 }\ \mathrm{and}\ 7.3\times { 10 }^{ -16 }\ \mathrm{erg\ {cm}^{-2}\ {s}^{-1}}$, and $6.0\times { 10 }^{ -16 }\ \mathrm{and}\ 3.0\times { 10 }^{ -15 }\ \mathrm{erg\ {cm}^{-2}\ {s}^{-1}}$, respectively. Upon correcting for systematic position offsets in each field, X-ray sources are cross-matched within a radius of 4$''$ of their $K_{\mathrm{s}}$-band counterparts. Of the 683 X-ray sources that overlap with the ZFOURGE fields, 545 ($\sim$80$\%$) are found within 1$''$ of a $K_{\mathrm{s}}$-band counterpart. A further 47 sources ($\sim$7$\%$) at $>$1$''$ are added after a visual inspection of both the X-ray and $K_{\mathrm{s}}$-band imaging confirmed no confusion from multiple sources within the matching radius. The remaining 91 sources yield no further matches with no visible counterparts identifiable. Considering this, a total of 592 X-ray counterparts are found in the ZFOURGE $K_{\mathrm{s}}$-band images ($\sim$87$\%$ of all overlapping X-ray sources), with 422 in CDFS, 93 in COSMOS, and 77 in UDS. 

\subsection{Far-infrared Data}

We make use of overlapping {\it Spitzer}/MIPS and {\it Herschel}/PACS FIR imaging. The data used in this study are from 24 and 160 $\mu$m photometry. We use imaging from the GOODS {\it Spitzer} Legacy program (PI: M. Dickinson) and GOODS-H \citep{Elbaz:2011ix} for the ZFOURGE-CDFS field, S-COSMOS Spitzer Legacy program (PI: D. Sanders) and CANDELS-H ({\color{blue}Inami et al.} {\color{blue}2015, in prep}) for the ZFOURGE-COSMOS field, and SpUDS {\it Spitzer} Legacy program (PI: J. Dunlop) and CANDELS-H for the ZFOURGE-UDS field. The median 1$\sigma$ flux uncertainties for each survey is $\sim$10 $\mu$Jy in COSMOS and UDS, and 3.9 $\mu$Jy in CDFS. Photometry from this data are produced using Multi-Resolution Object PHotometry oN Galaxy Observations (MOPHONGO) code written by I. Labb\'e \citep[for further details, see][]{Labbe:2006fo, Fumagalli:2014de, Whitaker:2014ko}.

\subsection{Photometric Redshifts, Rest-frame Colors, Stellar Masses and Star-Formation Rates}
\label{sec:sfrs}
The photometric redshifts and rest-frame colours of galaxies in ZFOURGE are calculated using the public SED-fitting code, \texttt{EAZY} \citep{Brammer:2008gn}. \texttt{EAZY} uses a default set of 5 templates generated from the \texttt{P\'{E}GASE} library \citep{Fioc:1997up}, plus an additional dust-reddened template from \citet{Maraston:2005er}. Linear combinations of these templates are fit to the observed $0.3-8$ $\mu$m photometry for estimating redshifts. Stellar masses are calculated by fitting \citet{Bruzual:2003ck} stellar population synthesis models using \texttt{FAST} \citep{Kriek:2009cs}, assuming solar metallicity, a \citet{Calzetti:2000iy} dust extinction law (with $A_V$ = $0-4$), a \citet{Chabrier:2003ki} initial mass function (IMF) and exponentially declining star-formation histories of the form SFR($t$) $\propto {e}^{{-t}/{\tau}}$, where $t$ is the time since the onset of star-formation and $\tau$ (varied over log[$\tau$/yr] $= 7-11$) modulates the declining function. Star-formation rates (SFRs) are calculated by considering both the rest-frame UV emission from massive unobscured stars and the re-radiated IR emission from dust obscured stars. The combined UV and IR luminosities ($L_{\mathrm{UV}}$ and $L_{\mathrm{IR}}$) are then converted to SFRs ($\Psi$) using the calibration from \citet{Bell:2005hs}, scaled to a \citet{Chabrier:2003ki} IMF:
\begin{equation}
\mathrm{\Psi}_{\mathrm{IR+UV}}\mathrm{[{\it M}_{\odot}} \ \mathrm{yr}^{-1}]=1.09\times{10}^{-10}(3.3{\it L}_{\mathrm{UV}} + {\it L}_{\mathrm{IR}})
\end{equation}

\noindent where ${L}_{\mathrm{UV}}$=$\nu{L}_{\nu,2800}$ is an estimate of the integrated 1216-3000\AA\ rest-frame UV luminosity, derived from \texttt{EAZY}, and ${L}_{\mathrm{IR}}$ is the bolometric 8-1000$\mu$m IR luminosity calculated from a luminosity-independent conversion \citep{Wuyts:2008hi, Wuyts:2011da} using PACS 160$\mu$m fluxes. For stacked measurements, we consider all sources, including those with zero or negative 160$\mu$m fluxes. This ensures our samples are not biased against quiescent galaxies or those with low SFRs. A comparison of our 160$\mu$m fluxes to that of the PACS Evolutionary Probe survey \citep{Lutz:2011ig} reveals good correspondence, with a median offset of $\Delta$mag $\sim0.20$. The quality of other derived galaxy parameters are explored in more depth in the ZFOURGE survey paper {\color{blue}Straatman et al.} ({\color{blue}2015, submitted}).

For all galaxies, whether active or inactive, we use `pure' galaxy templates in our SED fits, without consideration of an AGN component. Some studies adopt a single power-law template in an effort to decompose the combined SED into AGN and host galaxy components \citep[e.g.][]{Hao:2005fo, Bongiorno:2013jh, Rovilos:2014gh}. Though popular, it is unknown if such a broad approach would be effective on our diverse sample of AGN. We acknowledge potential contamination from AGN and adopt various tests to check for the effects on photometric redshifts (Section~\ref{sec:pz}) and other derived galaxy properties when presenting our results (Section~\ref{sec:con}).

\begin{figure}
\begin{center}
\includegraphics[width=\columnwidth]{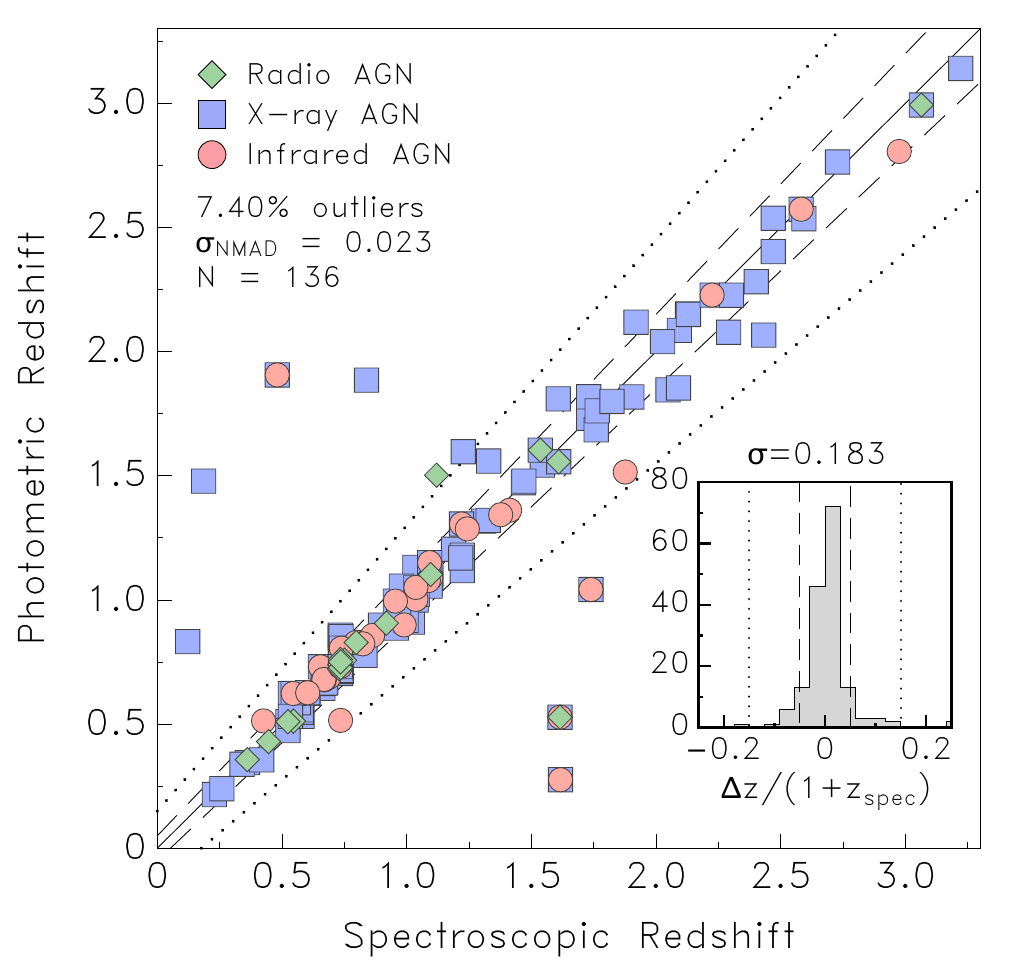}
\caption{Comparison of photometric and spectroscopic redshifts for our radio (green diamonds), X-ray (blue squares) and IR (red circles) AGN hosts. The solid line is the $z_{\mathrm{phot}} = z_{\mathrm{spec}}$ relation, the dashed lines are $z_{\mathrm{phot}} = 0.05 \pm (1+z_{\mathrm{spec}})$ and the dotted lines are $z_{\mathrm{phot}} = 0.15 \pm (1+z_{\mathrm{spec}})$. AGN Hosts outside of the dotted lines are defined as outliers. 
\label{fig:zvz}}
\end{center}
\end{figure}

\subsection{Reliability of AGN Photometric Redshifts}
\label{sec:pz}

AGN emission is known to complicate the computation of photometric redshifts \citep[e.g.][]{MacDonald:2010hz}, which can ultimately impact the derivation of rest frame colours and stellar population properties. In order to test the accuracy of our AGN sample (see Section~\ref{sec:data} for AGN classifications), we compare the sample's photometric redshifts from ZFOURGE to a secure sample of publicly available spectroscopic redshifts sourced from the compilation of the 3D-HST \citep{Skelton:2014do} and ZFIRE ({\color{blue}Nanayakkara et al.} {\color{blue}2015, submitted}) surveys. We use the Normalised Median Absolute Deviation (NMAD) to calculate scatter:
\begin{equation}
{\sigma}_{\mathrm{NMAD}}= 1.48\times \mathrm{median}\left( \frac { \left| \Delta z- \mathrm{median}(\Delta z) \right|  }{ 1+{ z }_{ \mathrm{spec} } }  \right) 
\end{equation}

\noindent where $\Delta z = z_{\mathrm{phot}} - z_{\mathrm{spec}}$. From the 500 AGN hosts identified in ZFOURGE, we find 136 cross-matches with reliable spectroscopic redshifts. Figure~\ref{fig:zvz} shows a relatively small number of AGN hosts with photometric redshifts very different from the spectroscopic value. These outliers (defined here to have ${ \left| \Delta z \right|  }/{ \left( 1+{ z }_{ { \mathrm{spec} } } \right)  }>0.15$) make up 7.40$\%$ of our sample and are subsequently ejected. Assuming the remainder of the AGN population has a similar outlier fraction, there is potential for an additional 27 AGN in our sample to have unreliable redshifts. Indeed, we visually inspect the SEDs of those AGN lacking a spectroscopic counterpart and manually eject 14 (3.85$\%$) with questionable fits. The accuracy of photometric redshifts for our AGN hosts is ${\sigma}_{\mathrm{NMAD}}$ = 0.023, which is only slightly higher than the general ZFOURGE population (${\sigma}_{\mathrm{NMAD}}$ = 0.018; \citealt{Tomczak:2014hw}). 

The strong correspondence between the photometric and spectroscopic redshifts in ZFOURGE is attributed to the efficient way the ZFOURGE medium-band filters trace the 4000\AA/Balmer breaks, which is driven by stellar light. Despite this, it remains possible that restframe optical AGN emission can increase the uncertainty of the photometric redshifts. For obscured (i.e. Type-2) AGN, several studies have demonstrated contamination to host galaxy properties is negligible \citep{Silverman:2009fz, Schawinski:2010bl, Xue:2010dk}. However, the AGN population in this work may also contain luminous, unobscured (i.e. Type-1) AGN, which may impact SED fits. To quantify how many of these might be in our sample, we search for objects (at all redshifts) with rest-frame $UVJ$ colours $\pm0.5$ mag around a SWIRE Type-1 QSO template \citep{Polletta:2007ha}. We find 23 sources ($\sim4\%$ of the parent AGN population) with these colours. A visual inspection of their SEDs reveal a sound fit to photometry, resulting in a photometric redshift with low error. Given this, we retain these sources in the parent AGN population. For our comparative analysis (Section~\ref{sec:uvj}), we select a mass and luminosity limited subsample from the parent AGN population (Section~\ref{sec:cuts}). Only 1 of the 23 sources with QSO colours is selected in this subsample.

\section{Multi-wavelength AGN Selection}
\label{sec:data}

The diverse and complex interactions between an AGN and its host galaxy make constructing a thorough and unbiased sample a formidable task. Variations in luminosity, morphology, orientation and dust obscuration dictate the need for a multi-wavelength, multi-technique approach. For example, while optical and X-ray selection techniques are both highly efficient, they break down when AGN hosts are heavily obscured by large amounts of gas and dust \citep[e.g.][]{Lacy:2006gh, Eckart:2009bd}. On the other hand, radio and IR selection techniques are relatively immune to dust extinction, but galaxies with copious amounts of star-formation can contaminate a sample \citep[e.g.][]{Condon:2002be, Donley:2005da}. In this section, we describe our approach to minimise such bias by constructing a robust AGN sample from multi-wavelength data. We restrict the sample to sources over $ z = 0.2-3.2$ with clean photometric detections in the ZFOURGE catalogues (e.g. near star and low-signal-to-noise flags). Further details of the ZFOURGE quality control flags will be presented in {\color{blue}Straatman et al.} {\color{blue}2015, submitted}. 

\begin{figure}
\begin{center}
\includegraphics[width=\columnwidth]{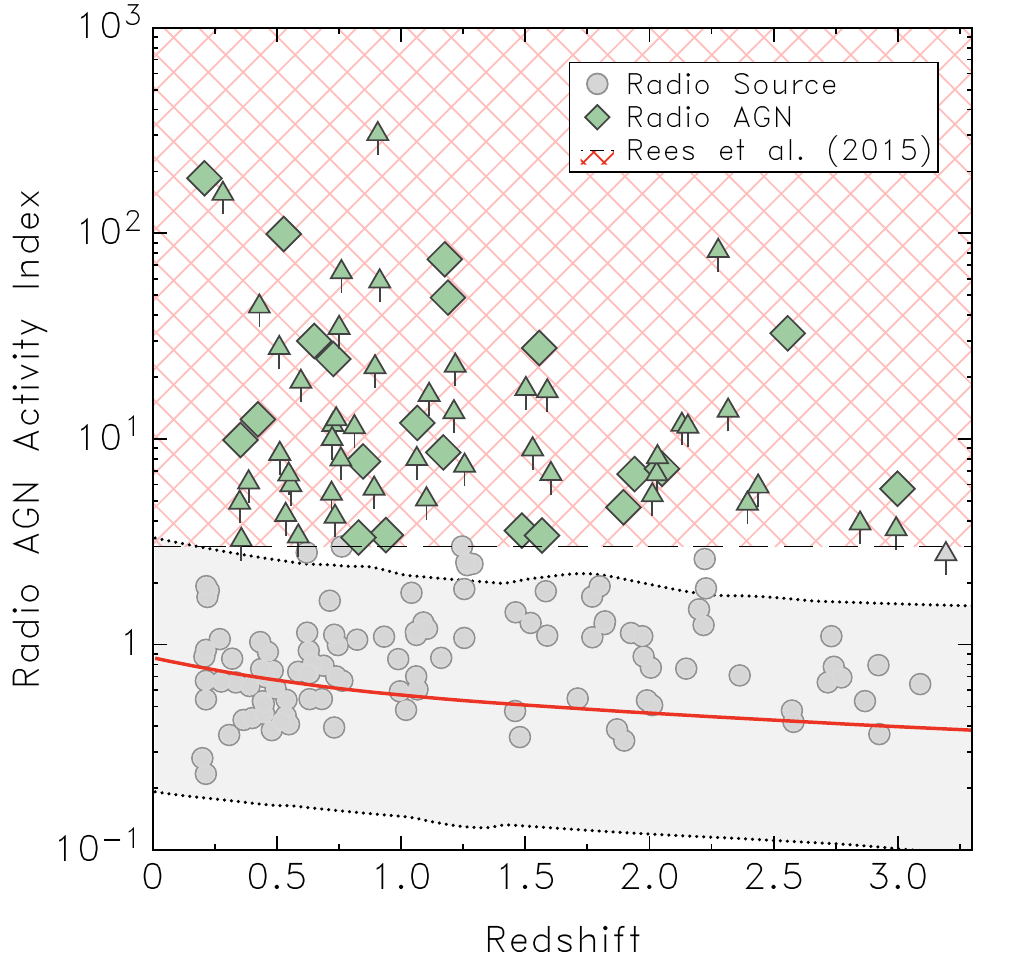}
\caption{The Radio AGN Activity Index (see equation \ref{eq:1}) for all radio sources in ZFOURGE. The evolution of the \citet{Wuyts:2008hi} average star-forming SED template, calculated from 160$\mu$m fluxes, is shown by the red line. The grey shaded region represents the $3\sigma$ 0.39 dex scatter found in the local radio-FIR correlation \citep{Moric:2010dg}. \citet{Rees:2015arXiv} adopt a conservative cut above this region (${ \mathrm{ \it{SFR} }_{ \mathrm{RAD} } }/{ \mathrm{ \it{SFR} }_{ \mathrm{IR+UV} }  }>3$; crosshatched region) to select radio AGN (green diamonds). Sources that lack a reliable ($>3\sigma$) 160$\mu$m detection are given $3\sigma$ limits (arrows).
\label{fig:radio}}
\end{center}
\end{figure}

\subsection{Radio AGN Selection}
\label{sec:ragn}

The accretion of material onto a supermassive black hole is known to produce nuclear radio emission, collimated into relativistic jets that propagate perpendicular to the plane of the accretion disc. While the detection of such radio-emitting jets unmistakably implies the presence of AGN, radio emission can also be caused by star-formation. As a result, radio detections at redshifts beyond the observable jet structure require alternative means for discriminating AGN hosts from inactive, star-forming galaxies. To achieve this, we use the Radio-AGN Activity Index of \citet{Rees:2015arXiv}. Briefly, this takes advantage of the tight correlation observed between a galaxy's radio (synchrotron) and IR (thermal) emissions \citep{Helou:1985be}, which \citet{Moric:2010dg} found holds for a diverse range of galaxies over a broad redshift. The exception was radio AGN, which presented a discernible offset from the correlation. The Radio-AGN Activity Index, which operates in SFR space, exploits this offset by assuming ${ \mathrm{ \it{SFR} }_{ \mathrm{RADIO} } } = { \mathrm{ \it{SFR} }_{ \mathrm{IR+UV} }  }$ if 100\% of the radio emission originates from star-formation. Sources with excess ${ \mathrm{ \it{SFR} }_{ \mathrm{RADIO} } }$ are identified as radio AGN:
\begin{equation}
\label{eq:1}
{ \mathrm{ \it{SFR} }_{ \mathrm{RADIO} } }/{ \mathrm{ \it{SFR} }_{ \mathrm{IR+UV} }  } = \text{Radio AGN Activity Index} >3
\end{equation}
The inclusion of UV emission accounts for the possibility of radio star-forming galaxies with low dust, which would otherwise produce an excess in ${\it SFR}_{\mathrm{RAD}}$ and be misclassified as radio AGN. To calculate radio SFRs, we first make use of the cross-matched photometric redshifts from ZFOURGE and apply radio $K$-corrections to estimate rest-frame radio luminosities using:
\begin{equation}
\quad { L }_{\mathrm{RADIO} }\mathrm{[W \ Hz^{-1}]}=4\pi { d }_{ l }^{ 2 }{ \left( 1+z \right)  }^{ -(\alpha +1) }{ f }_{\mathrm{RADIO} }
\end{equation}
\noindent where ${ d }_{ l }$ is the luminosity distance in cm, ${ f }_{\mathrm{RAD} }$ is the observed radio flux in W m$^{-2}$ Hz$^{-1}$, and $\alpha$ is the radio spectral index\footnote{The radio spectral index, $\alpha$ is defined from $S_{\nu} \propto \nu^{\alpha}$, where $S$ is the measured flux density and $\nu$ is the observer's frame frequency}, which we fix to $\alpha = -0.3$ as found in the \citet{Wuyts:2008hi} average star-forming SED template. While this spectral index is flatter than the standard $\alpha = -0.7$, it is adopted to ensure consistency with the \citet{Wuyts:2008hi} SED template, which is also used to derive IR SFRs. The difference between the two index values is one less source identified as a radio AGN under $\alpha = -0.7$. 

Using the rest-frame radio luminosities, radio SFRs are then calculated using the calibration from \citet{Bell:2003bj}, scaled to a \citet{Chabrier:2003ki} IMF:
\begin{equation}
\mathrm{\Psi}_{\mathrm{RADIO}}\mathrm{[{\it M}_{\odot}} \ \mathrm{yr}^{-1}]=3.18\times{10}^{-22}{\it L}_{\mathrm{RADIO}}
\end{equation}
As shown in Figure~\ref{fig:radio}, the Radio-AGN Activity Index leads to the identification of 67 radio sources dominated by AGN activity in ZFOURGE, with 20 in CDFS, 32 in COSMOS, and 15 in UDS.

\subsection{X-ray AGN Selection}

While radio surveys pioneered the way for AGN research \citep[e.g.][]{Baade:1954ig, Schmidt:1963kd, Schmidt:1964do}, the launch of {\it Chandra} and {\it XMM-Newton} heralded in a new era of sensitive, deep X-ray surveys, offering an effective alternative to select AGN. These surveys have found that X-ray emission from sources at high Galactic latitudes are predominantly AGN \citep[e.g.][]{Watson:2001jl} and routinely outshine the highest star-forming galaxies \cite[$\sim 10^{42}\ \mathrm{erg\ s^{-1}}$; e.g.][]{Moran:1999gr, Lira:2002jc}. While this provides an excellent discriminator for AGN selection, heavy obscuration by dense circumnuclear gas can prove problematic. One way to account for this is by examining the hardness ratio (HR) of a source, which is defined as the normalised difference of counts in the soft and hard X-ray bands, (hard - soft)$/$(hard + soft). The HR allows an estimate of absorption in the X-ray band, where obscured AGN are expected to exhibit a harder spectrum than unobscured AGN due to the absorption of soft X-rays by obscuring gas \citep{Szokoly:2004gz}. Considering this, we select X-ray AGN using both the X-ray luminosity and HR of a source. 

\begin{figure}
\begin{center}
\includegraphics[width=\columnwidth]{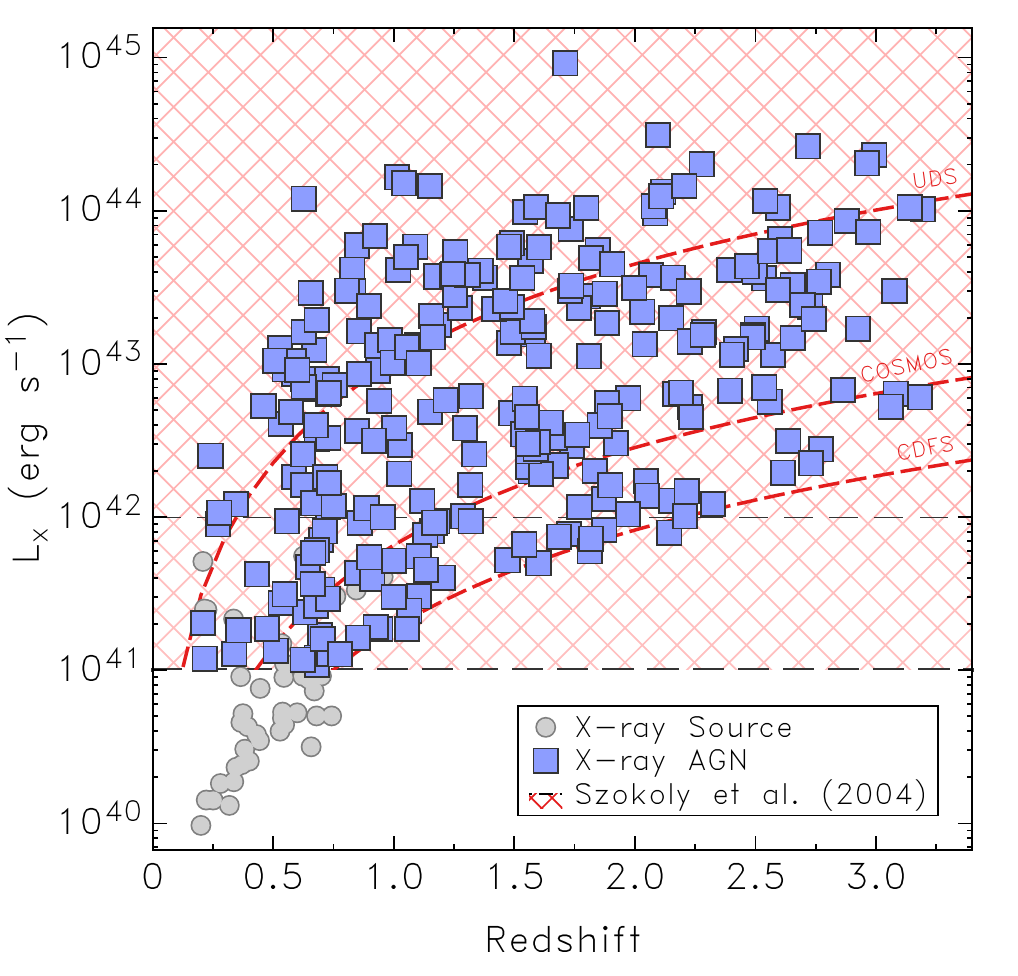}
\caption{X-ray rest-frame luminosity as a function of redshift for all X-ray sources in ZFOURGE. All sources above $10^{42}$ ergs s$^{-1}$  (upper crosshatched region; \citealt{Szokoly:2004gz}) are identified as AGN (blue squares), while only sources with a HR $>$ -0.2 down to $10^{41}$ ergs s$^{-1}$ (lower crosshatched region; \citealt{Szokoly:2004gz}) are identified as AGN. The approximate luminosity limits for each field are indicated by the red dashed curves.
\label{fig:xray}}
\end{center}
\end{figure}

We first start with the cross-matched photometric redshifts from ZFOURGE and apply X-ray $K$-corrections to estimate rest-frame luminosities using:
\begin{equation}
\quad { L }_{ \mathrm{X} }[\mathrm{erg \ s}^{-1}\mathrm{]} =4\pi { d }_{ l }^{ 2 }{ \left( 1+z \right)  }^{ \Gamma -2 }{ f }_{ x }
\end{equation}
\noindent where ${ d }_{ l }$ is the luminosity distance in cm, ${ f }_{ x }$ is the observed X-ray flux in erg cm$^{-2}$ s$^{-1}$, and $\Gamma$ is the photon index of the X-ray spectrum, which was fixed to a typical galaxy photon index\footnote{The photon index, $\Gamma$ is related to the number of incoming photons as a function of energy $\mathcal{E}$, $dN(\mathcal{E})/d\mathcal{E} \propto \mathcal{E}^{-\Gamma}$} of $\Gamma = 1.4$. For sources in the X11 catalogue, the intrinsic flux is derived from counts in the 0.5-8 keV full band, while for the E09 and U08 catalogues it is derived from the sum of the counts in the relevant bands over 0.5-10 keV. We adjusted flux values calculated in the E09 and U08 catalogues to align with the full bandpass values of the X11 catalogues (0.5-10 $\rightarrow$ 0.5-8 keV) assuming a power-law model of $\Gamma = 1.4$ (i.e. E09 and U08 fluxes are multiplied by a factor of 0.95). We then use the selection technique of \citet{Szokoly:2004gz} to select X-ray AGN:
\begin{align} 
{ L }_{ \mathrm{X} }\ge { 10 }^{ 41 }\mathrm{erg\ s}^{ -1 }\mathrm{\ \&\ HR} >-0.2 \nonumber \\
{ L }_{ \mathrm{X} }\ge { 10 }^{ 42 }\mathrm{erg\ s}^{ -1 }\mathrm{\ \&\ HR} \le -0.2
\end{align}
The luminosity threshold is lower for sources with a stronger HR on account of substantial absorption. In the absence of a HR measurement, we only select sources with ${ L }_{ x }\ge { 10 }^{ 42 }\mathrm{erg\ s}^{ -1 }$. As shown in Figure~\ref{fig:xray}, this approach leads to the identification of 270 X-ray sources dominated by AGN activity in ZFOURGE, with 187 in CDFS, 57 in COSMOS, and 26 in UDS. 

\begin{figure*}
\begin{center}
\includegraphics[width=\textwidth]{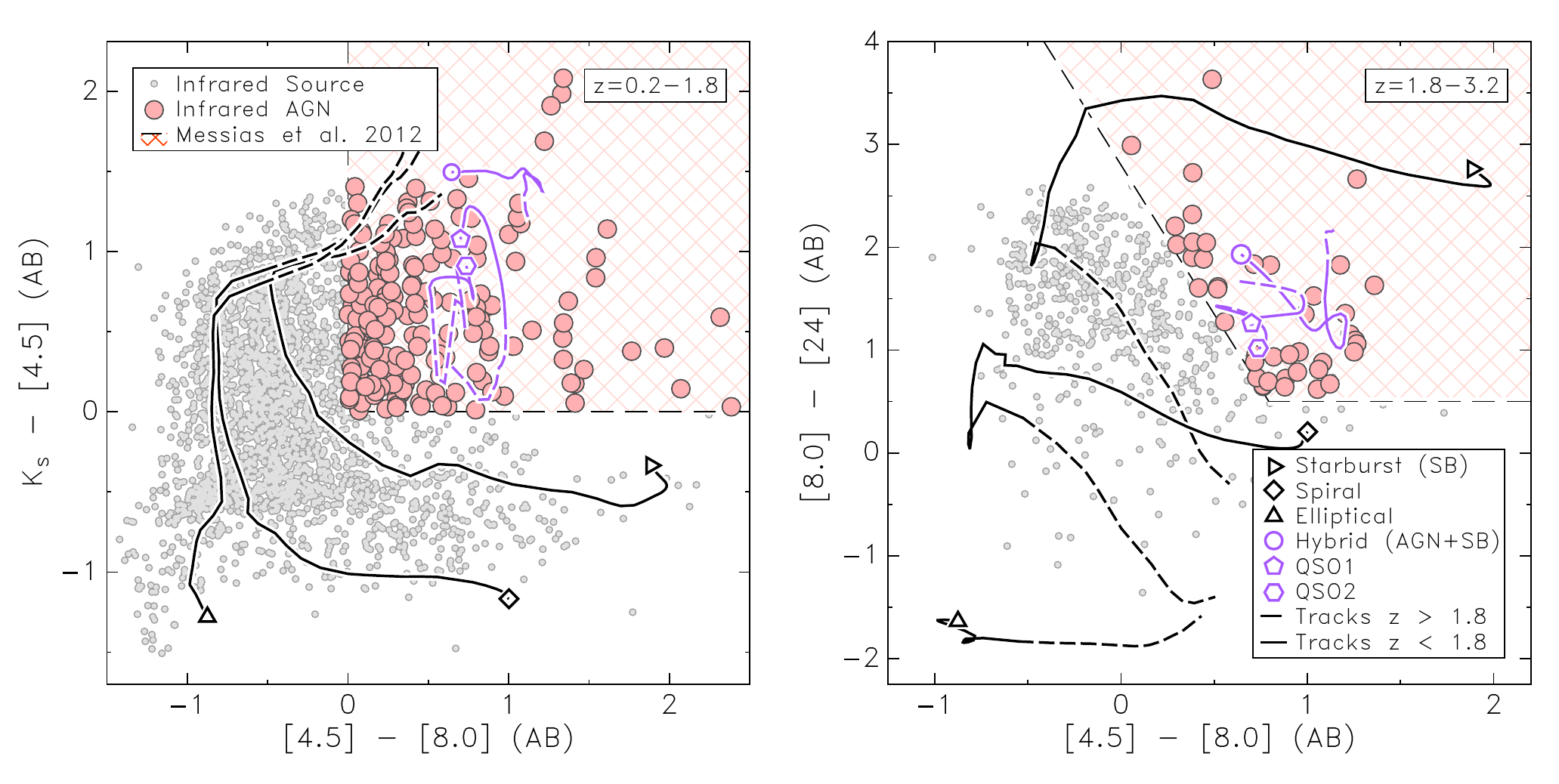}
\caption{The \citet{Messias:2012jq} $K_{\mathrm{s}}$ + IRAC (left) and IRAC + 24 $\mu$m (right) infrared colour-colour space for all IR sources within ZFOURGE. Sources that fall within the cross-hatched regions are considered AGN dominated (red circles). Overplotted are the redshift-dependent spectral tracks for a selection of active (hybrid, QSO1 and QSO2) and inactive (starburst, spiral and elliptical) galaxies from the Swire Templates \citep{Polletta:2007ha}. The dashed portion of the tracks represents $z = 0-1.8$ and the solid portion $z = 1.8-4$.}
\label{fig:ir}
\end{center}
\end{figure*}

\subsection{Infrared AGN Selection}
\label{sec:iragn}

Despite the efficiency of AGN selection in X-ray surveys, an imbalance in the cosmic X-ray background budget suggests an additional population of heavily obscured AGN are being missed \citep{Comastri:1995vs, Gilli:2001ge, Gilli:2007hb}. IR observations offer an effective way to identify these AGN by virtue of dust radiating the reprocessed nuclear emission in the mid-IR regime \citep{Sanders:1988fc, Sanders:1989bo}. Such emission is evident by the changing shape of a galaxy's SED, where an increase in AGN activity also leads to a dilution in the strength of Polycyclic Aromatic Hydrocarbon (PAH) emissions features formed by ultraviolet excitation typical in star-forming regions \citep{Brandl:2006fe}. The mid-IR is then dominated by the thermal continuum \citep[e.g.][]{Neugebauer:1979jq, Heisler:1999ho}. A number of IRAC colour-colour diagnostics have been designed to select AGN by taking advantage of this process \citep[e.g.][]{Lacy:2006gh, Stern:2005ey, Donley:2012ji}. The choice of diagnostic depends on the science being conducted as each has a particular level of completeness and reliability, with one often dominating in favour of the other \citep[e.g.][]{Barmby:2006ib, Donley:2007ci, Messias:2012jq}. Unfortunately, with increasing redshift, the IRAC bands begin to probe shorter rest-frame wavelengths and eventually trace the 1.6 $\mu$m stellar bump of a galaxy's SED, which can mimic the AGN thermal continuum. As a result, diagnostics limited to IRAC colours become ineffective at $z \gtrsim 2.5$ and rapidly introduce contaminants into the selection. \citet{Messias:2012jq} investigated this and found by extending the use of IRAC to additional wavebands, they could reliably select AGN over a broader redshift range. Specifically, the authors proposed two colour diagnostics, $K_{\mathrm{s}}$ + IRAC at lower redshifts ($z=0-2.5$) and IRAC + 24 $\mu$m at higher redshifts ($z=1-4$). We adopt these diagnostics with the added condition sources have a $5\sigma$ detection limit in all relevant bands to reduce scatter. To match the redshift bins used in our analysis (see Section~\ref{sec:zbin}), we select IR AGN based on the following constraints:
\begin{align}
z<1.8\begin{cases} { K }_{ s }-\left[ 4.5 \right] >0 \\ \left[ 4.5 \right] -\left[ 8.0 \right] >0 \end{cases}
\end{align}

\begin{align}
z>1.8\begin{cases} \left[ 8.0 \right] -\left[ 24 \right] >2.9\times \left( \left[ 4.5 \right] -\left[ 8.0 \right]  \right) +2.8 \\ \left[ 8.0 \right] -\left[ 24 \right] >0.5 \end{cases}
\end{align}
As shown in Figure~\ref{fig:ir}, this approach leads to the identification of 234 IR sources dominated by AGN activity in ZFOURGE, with 66 in CDFS, 50 in COSMOS, and 118 in UDS.

\subsection{Summary of AGN Samples}

\begin{figure*}
\begin{center}
\includegraphics[width=\textwidth]{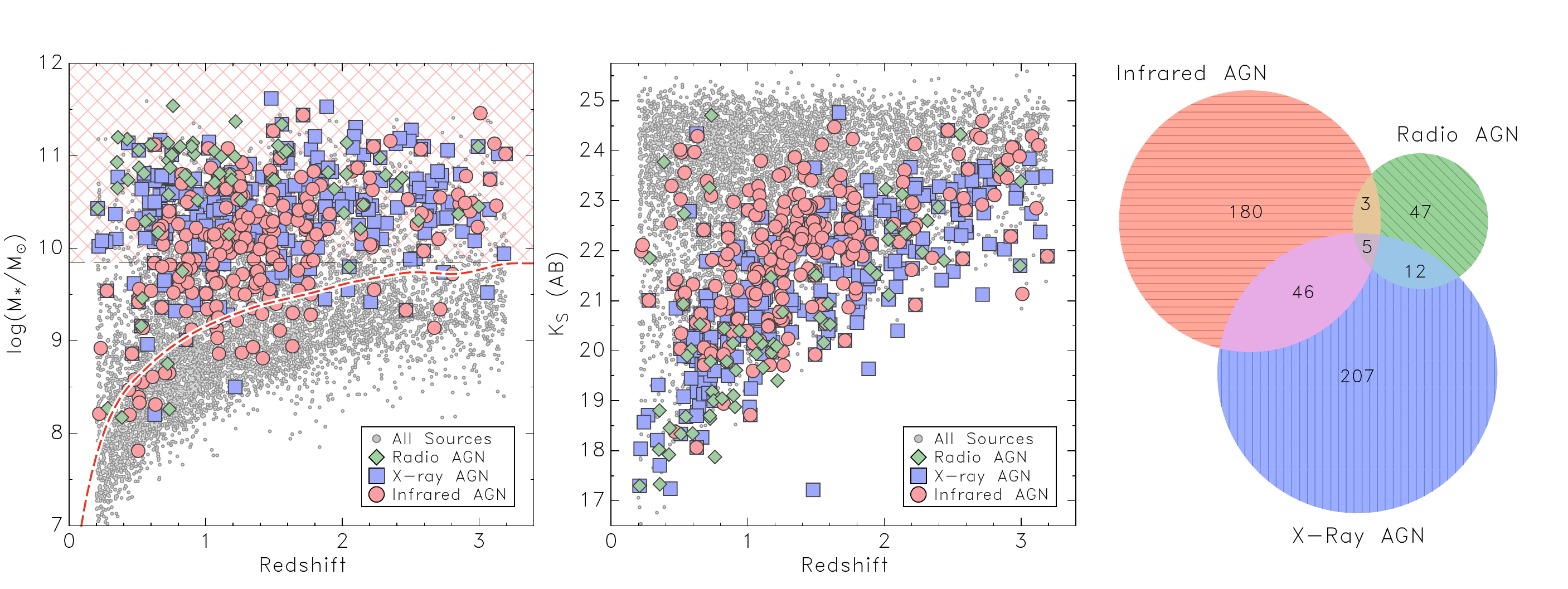}
\caption{Stellar mass (left) and $K_{\mathrm{s}}$-band magnitude (middle) as a function of redshift for our radio (green diamonds), X-ray (blue squares) and IR (red circles) AGN hosts, along with the parent sample from ZFOURGE (grey circles). The red dotted line in the left plot represents the 80\% mass-completeness limit in ZFOURGE, while the black dashed line is the stellar mass cut of log($M_{*}/M_{\odot}$) $\geq 9.75$. The Venn diagram (right) shows the relative number of AGN identified in radio (green), X-ray (blue) and IR (red) wavebands. The overlapping regions between samples correspond to the relative numbers selected in multiple wavebands. Note that these numbers correspond to the complete AGN candidate catalogue detailed in Section~\ref{sec:data}. For clarity, only ${1/3}^{\mathrm{rd}}$ of the parent sample is plotted. 
\label{fig:triple}}
\end{center}
\end{figure*}

We illustrate the relative size and overlap between the AGN samples in Figure~\ref{fig:triple} (right panel). Overlap arises from the complex and broad emission of AGN spectra and emphasises that our samples are not wholly independent and not simply relegated to either a radio, X-ray or IR selection bin. Despite this, the relative size of the overlap is comparable to previous studies that have performed multi-wavelgth AGN selection \citep{Hickox:2009in, Juneau:2013kz}. Like these studies, we find the overlap between radio and X-ray AGN hosts is low, while the overlap between IR and X-ray AGN hosts is significantly larger. Of the 500 AGN identified, 54 are found to overlap in one or more wavebands, with 5 identified in all three. For this work, overlapping AGN are treated as independent sources (i.e. 5 sources: a radio, X-ray and IR AGN) unless measurements are made on the combined AGN sample, in which case they are treated as a single source. We summarise the columns of the complete AGN dataset in Table~\ref{table:catalog}, which provides all host galaxy parameters used to select AGN in ZFOURGE. In Figure~\ref{fig:triple}, we display the stellar mass and $K_{\mathrm{s}}$-band distributions, along with the population numbers by way of a Venn diagram. This dataset acts as a complementary catalogue to the primary ZFOURGE catalogues. An amended version will be made available at \url{http://zfourge.tamu.edu} upon the full public release of ZFOURGE.

\section{Mass-Limited Sample}
\label{sec:cuts}
In this section, we extract AGN hosts from the catalogue of candidates selected in Section~\ref{sec:data} with the goal of constructing a mass-matched, inactive sample of galaxies (control sample) to compare star-formation activity between AGN hosts and inactive galaxies. Selection is based on redshift, stellar mass and luminosity limits, with the goal of minimising bias on host galaxy properties. Given the shallow X-ray and radio data used to select AGN hosts in ZFOURGE-UDS, this field will be excluded from the comparative analysis. 
\vspace{0.5cm}

\begin{table}
\setlength{\tabcolsep}{0.45em}
\centering
\caption{Luminosity limits of mass-limited AGN sample}
\begin{tabular}{l l l l c c c}
\hline\hline
\noalign{\vskip 1mm}  
Waveband & $L_{1.4\mathrm{GHz}}$ & $L_X$ & $L_{\mathrm{IR}}^{1}$ & $z_{\mathrm{min}}$ & $z_{\mathrm{max}}$ & $N_{\mathrm{AGN}}$$^2$ \\ [0.5ex] 
& (W Hz$^{-1}$) & (erg s$^{-1}$) & & &\\
\hline  
\noalign{\vskip 1mm}  
Radio & $1.0\times 10^{23}$ & - & -  & 0.2 & 0.8 & 10 \\[1ex]
& $6.0\times 10^{23}$ & - & -  & 0.8 & 1.8 & 11 \\[1ex]
& $1.9\times 10^{24}$ & - & -  & 1.8 & 3.2 & 5 \\[1ex]
X-ray & -  & $4.0\times 10^{41}$ & -  & 0.2 & 0.8 & 31 \\ [1ex] 
& -  & $2.0\times 10^{42}$ & -  & 0.8 & 1.8 & 60 \\[1ex]
& -  & $7.0\times 10^{42}$ & -  &  1.8 & 3.2 & 50 \\[1ex]
Infrared & - & - & $6.0\times 10^{27}$ & 0.2 & 0.8 & 7 \\[1ex]
& - & - & $3.0\times 10^{28}$ & 0.8 & 1.8 & 39 \\[1ex]
& - & - & $1.0\times 10^{27}$ & 1.8 & 3.2 & 22 \\[1ex]
\hline
\end{tabular}
  \\[0.2cm] 
  \raggedright \footnotesize $^1$ $L_{\mathrm{IR}}$ = $L_{\mathrm{8\micron}}$ at $z = 0.2-1.8$ and $L_{\mathrm{24\micron}}$ at $z = 1.8-3.2$\\
  $^2$ Number of AGN hosts within the specified limits\\
\label{table:limits}
\end{table}

\subsection{Redshift, Mass and Luminosity Cuts}
\label{sec:zbin}
To overcome the potential bias associated with $K_{\mathrm{s}}$-band selected galaxies, we limit our sample of AGN hosts to a stellar-mass cut of log($M_{*}/M_{\odot}$) $\geq 9.75$, which sits above the 80\% completeness limit of ZFOURGE \citep{Papovich:2015kn}, as shown in Figure~\ref{fig:triple} (left panel). We apply further restrictions by splitting the AGN sample into three redshift bins of $z =$ [0.2-0.8], [0.8-1.8], [1.8-3.2], each with varying luminosity limits based on the luminosity thresholds of their respective wavebands (i.e. $L_{1.4\mathrm{GHz}}$, $L_X$ and $L_{\mathrm{IR}}$). These limits are summarised in Table~\ref{table:limits} and while they reduce AGN numbers and restrict comparison across redshifts, they minimise potential luminosity biases by ensuring a consistent luminosity-completeness within each redshift bin.

\subsection{Control Sample of inactive Galaxies}

Tight correlations exist between the physical properties of galaxies and their stellar mass (e.g. \citeauthor{Tremonti:2004ed}, \citeyear{Tremonti:2004ed}, mass-metallicty and \citeauthor{Noeske:2007ja}, \citeyear{Noeske:2007ja}, mass-star-formation rate). This makes constructing a mass-matched control sample of inactive galaxies an essential component for our comparative analysis. Without this consideration, even a mass-limited sample would be dominated by galaxies just above the mass threshold, potentially biasing any comparison. 
We construct our mass-matched control sample by binning inactive galaxies into narrow mass intervals of $\Delta M _*$ = 0.2 dex. 

\begin{figure*}
\begin{center}
\includegraphics[width=\textwidth]{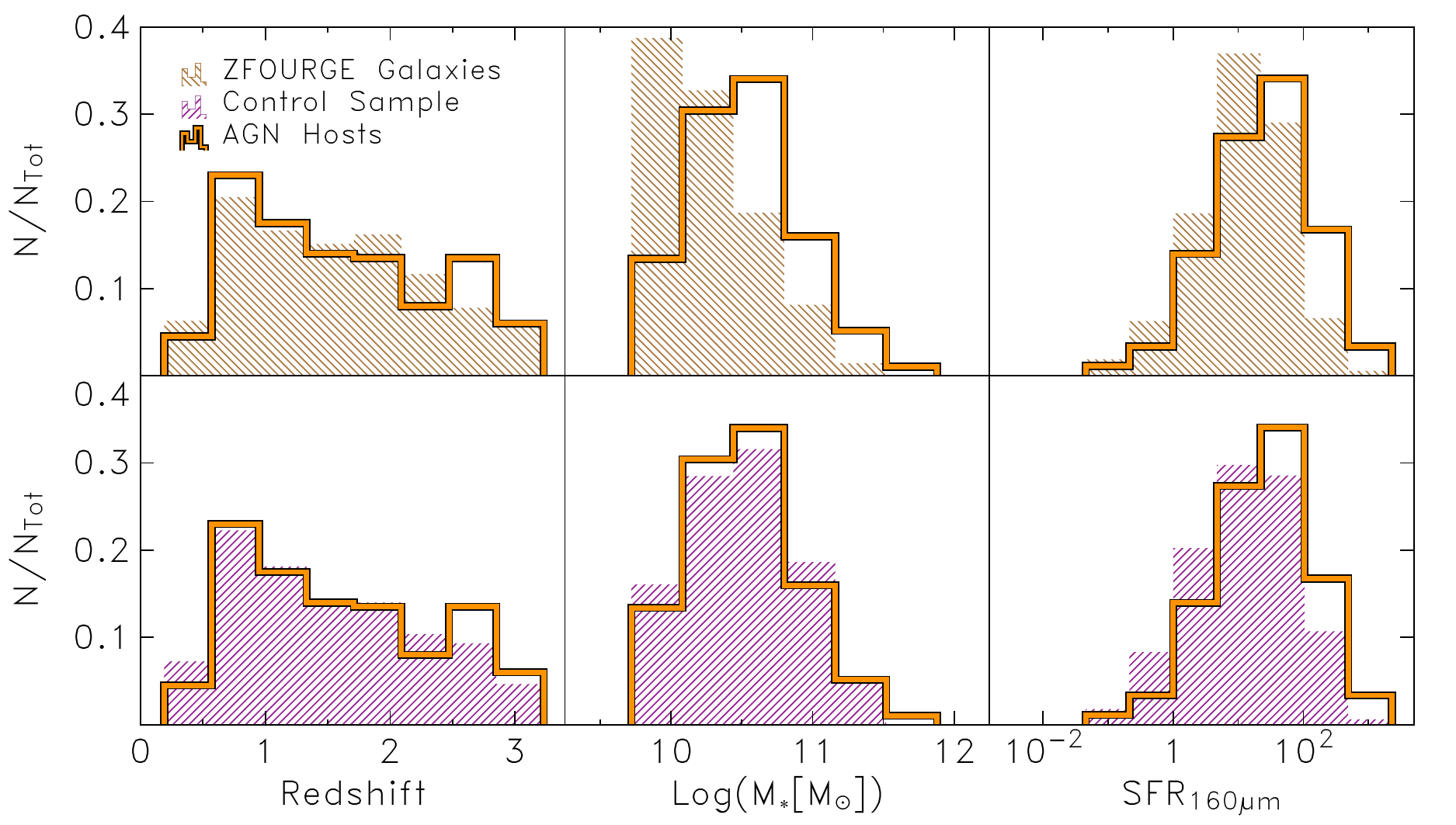}
\caption{The redshift (left), stellar mass (middle) and SFR (right, limited to positive fluxes) distributions for the parent population of galaxies (top row, hatched), control sample of inactive galaxies (bottom row, hatched), and luminosity limited AGN hosts (solid orange line) in ZFOURGE.}
\label{fig:histogram}
\end{center}
\end{figure*}

For each AGN host, we randomly select an inactive galaxy from the same redshift bin ($z =$ [0.2-0.8], [0.8-1.8] or [1.8-3.2]) and of similar mass, within $\Delta M _*$. For example, a $z=0.74$ radio AGN host with log($M_{*}/M_{\odot}$) $=10.87$ has 112 inactive analogues from which to draw from. We then calculate and record a mean value for various physical properties of the selected control inactive galaxy (i.e. rest-frame colour, stellar mass and star-formation rate) and repeat for the next AGN host until we have a control sample with the same number of galaxies as the AGN sample. We generate 100 such independent control samples, which we use to compute a final mean control value for each physical property. The distribution of various physical properties for the mass-limited sample of AGN and control sample of inactive galaxies is shown in Figure~\ref{fig:histogram}.

\section{Results}
\subsection{Comparison of Rest-frame Colours}
\label{sec:uvj}
\begin{figure*}
\begin{center}
\includegraphics[width=\textwidth]{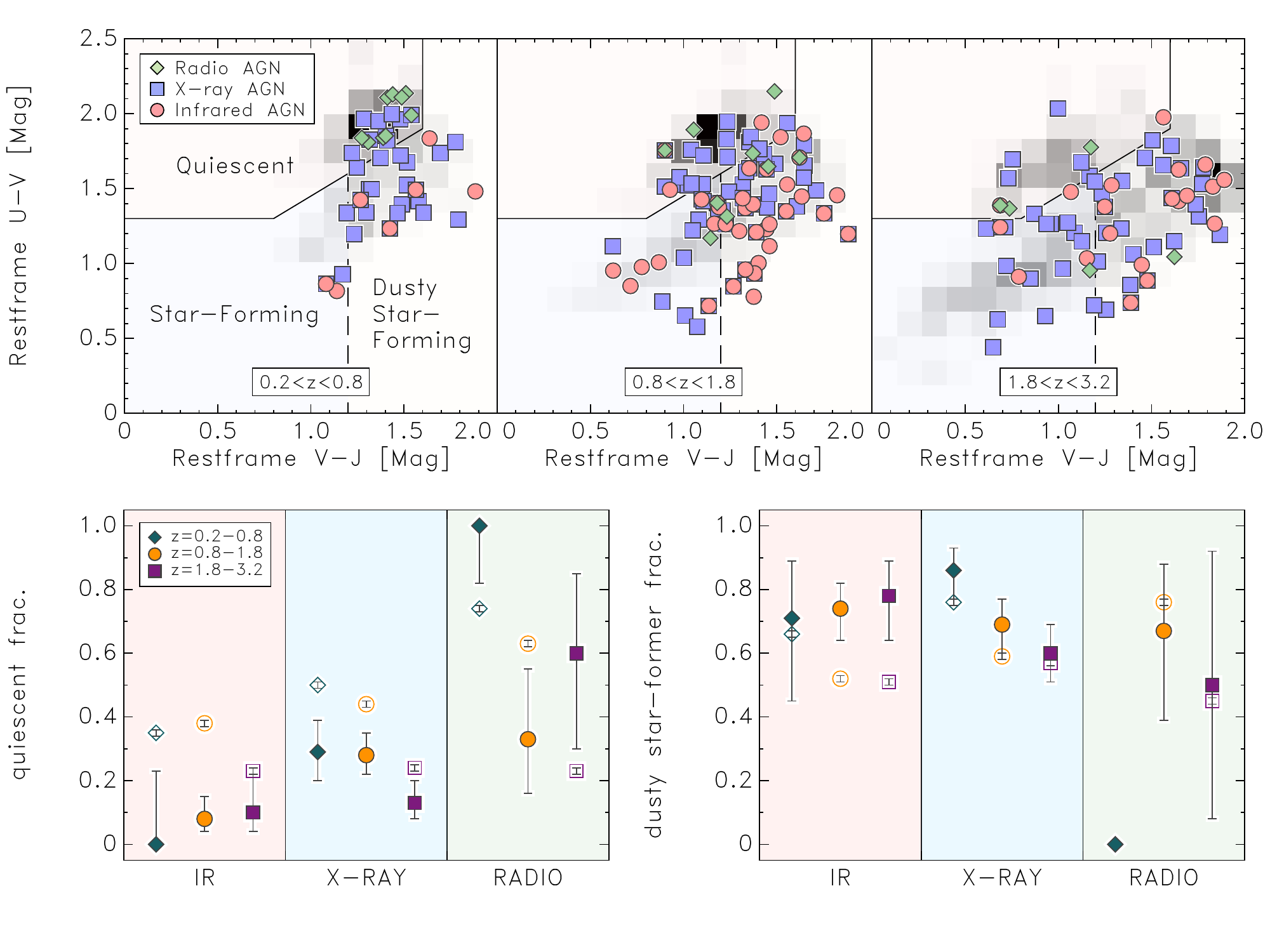}
\caption{(Top) The rest-frame $UVJ$ colour classification of galaxies in bins of redshift ($z = 0.2-0.8$; left, $z = 0.8-1.8$; middle and $z = 1.8-3.2$; right). The points represent the mass-limited (log($M_{*}/M_{\odot}$) $\geq 9.75$) AGN hosts selected via radio (green diamonds), X-ray (blue squares) and IR (red circles) techniques. A representation of the control sample is shown by the grey-scale density plot in each panel. The solid line divides the population into quiescent and star-forming hosts, while the dashed line further divides the star-forming population into dusty and non-dusty galaxies. (Lower-left) The quiescent fraction (${N}_{\mathrm{q}}/({N}_{\mathrm{q}}+{N}_{\mathrm{sf}})$) and (lower-right) dusty star-former fraction (${N}_{\mathrm{dusty}}/{N}_{\mathrm{sf}}$) for the mass-limited AGN hosts (closed markers) and the control sample (open markers) at $z = 0.2-0.8$ (diamond markers), $z = 0.8-1.8$ (circle markers) and $z = 1.8-3.2$ (square markers). Values are derived from the $UVJ$ colour classification. Vertical error bars indicate the $1\sigma$ Clopper-Pearson confidence intervals. Unless shown, error bars are smaller than the plotting symbols for the control sample.}
\label{fig:uvj}
\end{center}
\end{figure*}

Examining the rest-frame $UVJ$ colours of galaxies has become a common approach to distinguish a quiescent population from a star-forming one, including those exhibiting heavy extinction \citep[e.g.][]{Labbe:2005dp, Wuyts:2007hs, Williams:2009hn}. Referring to the top panel in Figure~\ref{fig:uvj}, quiescent galaxies occupy the upper left region, delimited by the vertices ($V-J, U-V$) = ($-\infty,1.3$), ($0.85,1.3$), ($1.6,1.95$), ($1.6,+\infty$), while the vertical dashed-line ($V-J = 1.2$) separates non-dusty (lower left) from dusty star-forming galaxies \citep{Spitler:2014ey}. 

Within this figure, we examine the $UVJ$ colour space of our mass-limited AGN hosts and control sample of inactive galaxies. In the lowest redshift bin ($z = 0.2-0.8$), we find the $UVJ$ colours of each subsample of AGN, identified in radio, X-ray or IR, to be consistent with a distinct galaxy population. IR AGN are found exclusively in star-forming galaxies, radio AGN in quiescent galaxies, and X-ray AGN in both quiescent ($29.0\%\pm8.2\%$) and star-forming hosts. However, at higher redshifts ($z>0.8$), the trend weakens and the distribution of $UVJ$ colours scatter to the point where all AGN are predominantly found in the colour space of star-forming hosts (radio AGN; $57.1\%\pm13.2\%$, X-ray AGN; $79.0\%\pm4.1\%$, IR AGN; $91.2\%\pm3.8\%$), mirroring the behaviour of the control sample of galaxies. 

When comparing the distribution of $UVJ$ colours between AGN hosts and the control sample, the two are found to be qualitatively similar at all redshifts, with slight differences in the peak of their distributions. To accentuate these differences and examine their impact, we compare the quiescent fraction (${f}_{\mathrm{q}}={N}_{\mathrm{q}}/({N}_{\mathrm{q}}+{N}_{\mathrm{sf}})$) and dusty star-former fraction (${f}_{\mathrm{dusty}}={N}_{\mathrm{dusty}}/{N}_{\mathrm{sf}}$) of both samples in Figure~\ref{fig:uvj} (lower panels). While low numbers in the radio AGN population hinder the ability to produce statistically significant results, offsets are observed between the IR and X-ray AGN hosts and their respective control samples. For both populations over all redshifts, the dusty fraction is found to be slightly elevated over the control samples, while the quiescent fraction is lower.

Together, all panels in Figure~\ref{fig:uvj} reveal no significant differences between the $UVJ$ colours of our AGN and control samples, with the exception that the AGN hosts tend to be more dusty and hosted in a lower fraction of quiescent galaxies. In the following section, we explore these results in further detail by quantitatively gauging the difference in star-formation activity between both samples.

\begin{figure}
\begin{center}
\includegraphics[width=\columnwidth]{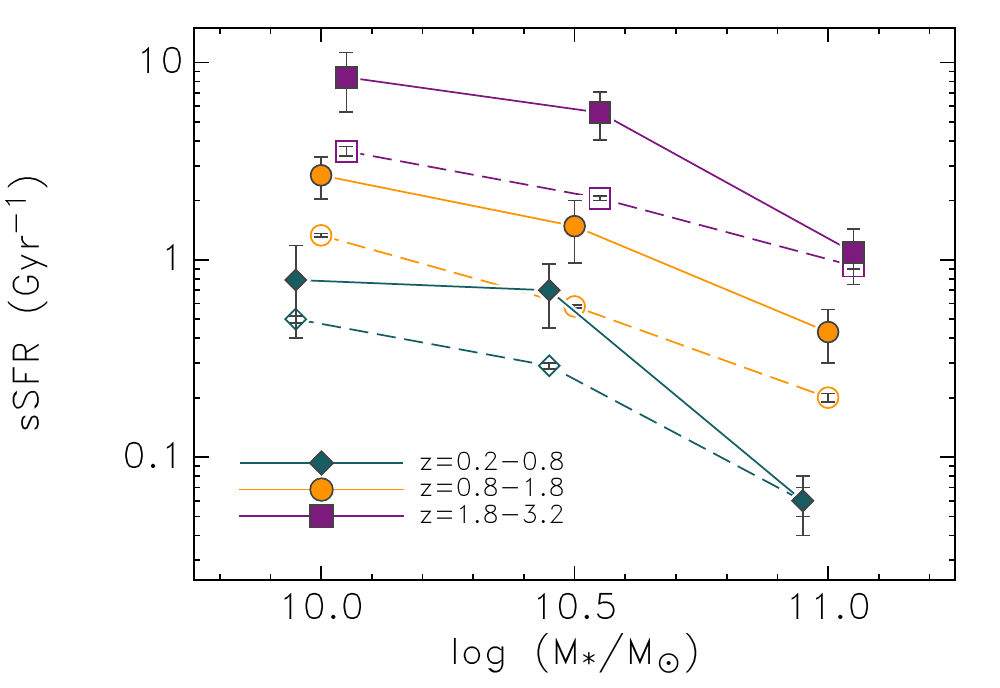}
\caption{The mean specific star-formation rate (SFR/M$_*$) as a function of stellar mass for the mass-limited (log($M_{*}/M_{\odot}$) $\geq 9.75$) AGN hosts (solid lines) and the control sample (dashed lines) at $z = 0.2-0.8$ (diamond markers), $z = 0.8-1.8$ (circle markers) and $z = 1.8-3.2$ (square markers). Error bars indicate the 68$\%$ confidence intervals evaluated from a bootstrap analysis. Unless shown, error bars are smaller than the plotting symbols for the control sample. The stellar mass of markers are offset for better visibility. With the exception of the highest mass bins at low and high redshifts, AGN hosts show an elevated level of star-formation activity with respect to the control sample of inactive galaxies.}
\label{fig:sfd}
\end{center}
\end{figure}

\subsection{Comparison of Star-Formation Activity}
\label{sec:sfa}
We now focus on the star-formation activity in our mass-limited AGN hosts and how they compare to the control sample of inactive galaxies. We use specific star-formation rate (sSFR) as a measure of the relative strength of star-formation activity, which is a galaxy's SFR normalised by the mass of its stars ($\mathrm{\Psi}_{\mathrm{IR+UV}}$/M$_*$). In Figure~\ref{fig:sfd}, we show the mean sSFR against stellar mass for our AGN hosts and control sample in bins of redshift. The mean sSFR is found to decrease with increasing stellar mass for all sources, with slight offsets observed between the AGN hosts and control sample. AGN hosts exhibit an elevation over the control sample, with an average logarithmic offset (linear average of the difference between the logarithmic sSFRs) for the combined mass bins of $0.26\pm0.14$ dex at $z = 0.2-0.8$, $0.37\pm0.10$ dex at $z = 0.8-1.8$, and $0.38\pm0.10$ dex at $z = 1.8-3.2$ (see Table~\ref{tab:ssfr} for more details).

To better understand the source of this offset, we split the AGN population by detection technique (i.e. radio, X-ray and infrared). In Figure~\ref{fig:sfa}, the mean sSFR of each subsample of AGN hosts, along with their respective control sample is shown. It can be seen that each subsample exhibits an elevated level of sSFR over their control samples, with the exception of low-redshift radio AGN hosts. For each subsample, the elevation is found to increase with redshift. This elevation is found to be consistently high and significant for IR AGN ($0.48\pm0.21$ dex; $z = 0.2-0.8$, $0.50\pm0.12$ dex; $z = 0.8-1.8$, $0.72\pm0.13$ dex; $z = 1.8-3.2$), but lower and insignificant for X-ray AGN ($0.15\pm0.13$ dex; $z = 0.2-0.8$, $0.21\pm0.13$ dex; $z = 0.8-1.8$, $0.25\pm0.16$ dex; $z = 1.8-3.2$). While high redshift radio AGN hosts ($z>0.8$) also present an elevated sSFR over the control sample, low number statistics impact its significance ($-0.53\pm0.20$ dex; $z = 0.2-0.8$, $0.57\pm0.20$ dex; $z = 0.8-1.8$, $0.55\pm0.32$ dex; $z = 1.8-3.2$ (see Table~\ref{tab:split} for more details). 

\begin{table*} 
\setlength{\tabcolsep}{1.3em}
\centering 
\caption{Mean sSFR (Gyr$^{-1}$) values by redshift bin (rows) for AGN hosts and the control in bins of stellar mass (cols)} 
\def\arraystretch{1}
\begin{tabular}{l c c c c c c} 
\hline \hline
& \multicolumn{3}{c}{AGN Hosts} &
\multicolumn{3}{c}{Control Sample} \\ 
\cmidrule(l){2-7} 
Redshift & $10^{9.75-10.25}M_{\odot}$ & $10^{10.25-10.75}M_{\odot}$ & $10^{10.75-11.25}M_{\odot}$ & $10^{9.75-10.25}M_{\odot}$ & $10^{10.25-10.75}M_{\odot}$ & $10^{10.75-11.25}M_{\odot}$\\ 
\midrule 
{\it z} = $0.2 - 0.8$ & 0.79$\pm 0.39$ & 0.70$\pm 0.25$ & 0.06$\pm 0.02$ & 0.50$\pm 0.02$ & 0.29$\pm 0.01$ & 0.06$\pm 0.01$\\ 
& $\sigma = 0.19$ & $\sigma = 0.21$ & $\sigma = 0.01$ & $\sigma = 0.11$ & $\sigma = 0.06$ & $\sigma = 0.02$\\
{\it z} = $0.8 - 1.5$ & 2.68$\pm 0.64$ & 1.48$\pm 0.52$ & 0.43$\pm 0.13$ & 1.33$\pm 0.03$ & 0.58$\pm 0.01$ & 0.20$\pm 0.01$\\ 
& $\sigma = 1.39$ & $\sigma = 0.60$ & $\sigma = 0.14$ & $\sigma = 0.28$ & $\sigma = 0.09$ & $\sigma = 0.05$\\
{\it z} = $1.5 - 2.5$ & 8.42$\pm 2.83$ & 5.57$\pm 1.54$ & 1.09$\pm 0.34$ & 3.55$\pm 0.20$ & 2.05$\pm 0.05$ & 0.93$\pm 0.03$\\ 
& $\sigma = 2.52$ & $\sigma = 2.40$ & $\sigma = 0.49$ & $\sigma = 1.16$ & $\sigma = 0.39$ & $\sigma = 0.19$\\
\hline
\end{tabular}
\raggedright \footnotesize \textbf{Notes.} Uncertainties are from a bootstrap analysis. Dispersions around the mean ($\sigma$) on quantities are median absolute deviations (MAD).\\
\label{tab:ssfr} 
\end{table*}

\subsection{AGN Contamination}
\label{sec:con}
As discussed in Section~\ref{sec:sfrs}, there is potential for AGN contamination to impact galaxy properties used in this analysis. The first is our SFRs, which are derived from a combination of UV and IR luminosities and may contain a mixed contribution of light from stars and AGN. We first examine the impact to the UV by removing the UV contribution to the SFRs of the AGN sample and recalculating our results. We find the offsets increase an average of 0.01 dex in each redshift bin, suggesting there is negligible impact from AGN contamination in the UV regime. If we assume contamination to the IR regime wholly explains the elevation of star-formation activity observed in our AGN sample, the contribution from AGN emission would need to be in excess of $\sim 25\%$. However, the FIR regime is thought to be mostly immune to the effects of AGN \citep[e.g.][]{Netzer:2007bs, Mullaney:2011ce}, which is the primary motivation for employing PACs-based SFRs.

The other potential impact is AGN contamination to stellar masses. \citet{Ciesla:2015vi} inspected this by omitting an AGN component while performing SED fitting on a range of Type-I, intermediate type, and Type-II AGN and comparing the measured stellar mass to the true value. Their results showed contamination from a Type-I AGN can lead to an overestimation in mass by as much as $150\%$. The contamination from intermediate and Type-II, believed to dominate the sample in this study, was overestimated by $\sim50\%$. We examine the most extreme of these cases ($150\%$ overestimation) and how it impacts our results. We first reduce the mass of our AGN population and then re-sample our mass-matched control sample of inactive galaxies. We find the total average logarithmic offset between active and inactive galaxies to decrease from $0.34\pm0.07$ dex to $0.25\pm0.07$ dex. Since the masses of AGN hosts are only ever overestimated by the SED fits, any sSFR discrepancy is considered to be a minor effect, if this systematic is present.

\begin{table*} 
\setlength{\tabcolsep}{1.35em}
\centering 
\caption{Mean sSFR (Gyr$^{-1}$) values by redshift bin (rows) for IR, X-ray and radio AGN hosts and their respective control samples (cols)} 
\def\arraystretch{1}
\begin{tabular}{l c c c c c c} 
\hline \hline
& \multicolumn{3}{c}{AGN Hosts} &
\multicolumn{3}{c}{Control Sample} \\ 
\cmidrule(l){2-7} 
Redshift & IR Active & X-ray Active & Radio Active & IR inactive & X-ray inactive & Radio inactive\\ 
\midrule 
{\it z} = $0.2 - 0.8$ & 1.57$\pm 0.68$ & 0.43$\pm 0.12$ & 0.04$\pm 0.01$ & 0.52$\pm 0.02$ & 0.31$\pm 0.01$ & 0.12$\pm 0.01$\\ 
& $\sigma = 0.19$ & $\sigma = 0.14$ & $\sigma = 0.01$ & $\sigma = 0.16$ & $\sigma = 0.05$ & $\sigma = 0.05$\\
{\it z} = $0.8 - 1.5$ & 2.74$\pm 0.64$ & 1.24$\pm 0.34$ & 1.18$\pm 0.49$ & 0.87$\pm 0.02$ & 0.77$\pm 0.02$ & 0.32$\pm 0.01$\\ 
& $\sigma = 1.18$ & $\sigma = 0.40$ & $\sigma = 0.39$ & $\sigma = 0.15$ & $\sigma = 0.10$ & $\sigma = 0.10$\\
{\it z} = $1.5 - 2.5$ & 10.56$\pm 2.75$ & 3.45$\pm 1.20$ & 8.23$\pm 5.37$ & 2.02$\pm 0.08$ & 1.96$\pm 0.05$ & 2.30$\pm 0.19$\\ 
& $\sigma = 3.15$ & $\sigma = 1.04$ & $\sigma = 2.48$ & $\sigma = 0.52$ & $\sigma = 0.36$ & $\sigma = 1.02$\\
\hline
\end{tabular}\\
\raggedright \footnotesize \textbf{Notes.} Uncertainties are from a bootstrap analysis. Dispersions around the mean ($\sigma$) on quantities are median absolute deviations (MAD).\\
\label{tab:split} 
\end{table*}

\section{Discussion}
\label{sec:dis}
While numerous studies have examined the difference between star-formation activity in AGN hosts and inactive galaxies, their results tend to be conflicting. Earlier studies, which were often limited to low-redshifts, low sample sizes, and no control or crudely matched comparison samples, predominantly found suppressed star-formation activity in AGN hosts \citep[e.g.][]{Ho:2005en, Kim:2006kw, Salim:2007hq}. However, with improved selection techniques and deeper observations, recent findings have found their star-formation activity is more similar or even elevated over inactive galaxies \citep{Xue:2010dk,  Mullaney:2011ce, Juneau:2013kz}. By implementing multiple AGN selection techniques and pushing to higher redshifts with deep multi-wavelength data, the present work supports the latter. 

Predominantly, we find that all AGN hosts exhibit a slight elevation in star-formation activity over inactive galaxies. This elevation is consistent across all redshifts, but less pronounced at high stellar mass. The exception to this elevation is low-redshift, radio AGN. As seen in Figure~\ref{fig:uvj}, this population is found to be exclusively hosted by quenched galaxies, which exhibit a lower level of star-formation activity than their mass-matched, inactive counterparts (see Table~\ref{tab:split}). For early studies, limited to low-redshifts, this was well established \citep[e.g.][]{Matthews:1964jz}, and possibly led to an early perception that AGN are associated with quenched, elliptical galaxies. Unlike infrared and high redshift ($z > 0.8$) radio AGN hosts, which exhibit a strong elevation in star-formation over their respective control samples, we find the offset for galaxies hosting X-ray AGN to be only marginal. Recent studies, while different in their approach, tend to find similar results. For example, in \citet{Bongiorno:2013jh}, the authors found their sample of type-II AGN to have, on average, the same or slightly lower SFRs than inactive galaxies of the same mass and redshift. \citet{Mullaney:2015hb} find the same, but compare their AGN sample to a main-sequence of star-forming galaxies. While we find slightly higher star-formation in X-ray AGN hosts, this can possibly be explained away by our different approach and selection effects (i.e., our star-formation estimates, mass, luminosity and redshift cuts). That being said, the overarching theme is consistent between all of these recent studies - the star-formation activity of X-ray AGN hosts is mostly consistent with normal galaxies.

While the elevated levels of star-formation in X-ray AGN is at best marginal, the offset between IR AGN and the control sample is explicit. The mean sSFR for IR AGN hosts is found to be as much as $\sim$5 times higher, suggesting there exists a stronger link between IR AGN and its host, than in other types of AGN. Such an analysis has not been accomplished before at high-redshifts due to concerns of AGN contamination in sSFR estimates. However, we mitigate against this effect by employing 160$\mu$m derived SFRs, which is believed to be predominantly free from AGN activity.

\begin{figure}
\begin{center}
\includegraphics[width=\columnwidth]{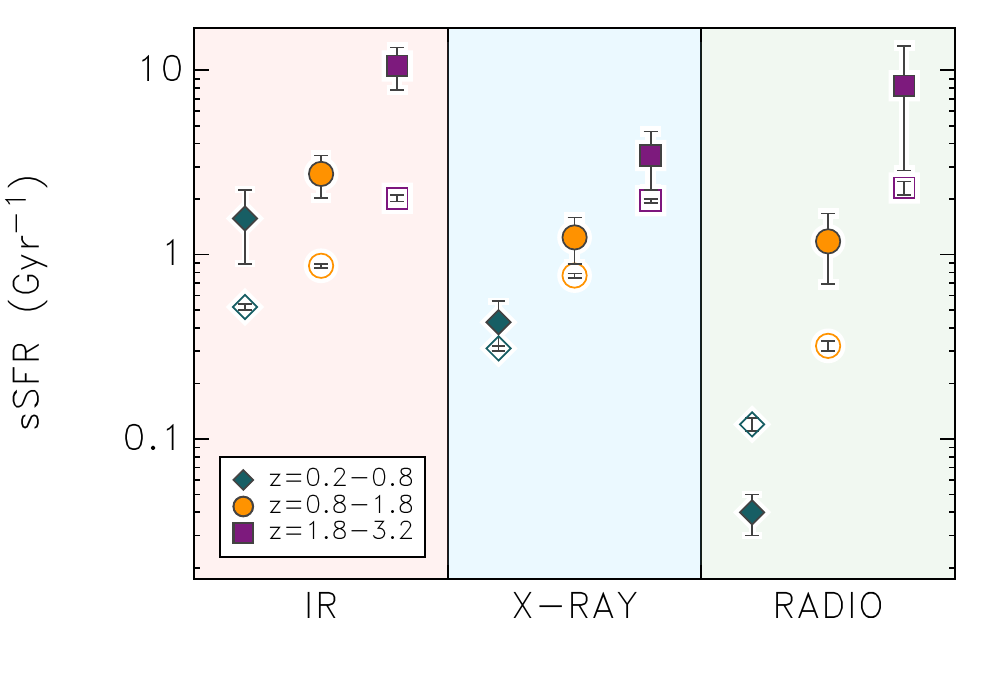}
\caption{The mean specific star-formation rate, split by AGN class (IR, X-ray and radio) for the mass-limited (log($M_{*}/M_{\odot}$) $\geq 9.75$) AGN hosts (closed markers) and the control sample (open markers) at $z = 0.2-0.8$ (diamond markers), $z = 0.8-1.8$ (circle markers) and $z = 1.8-3.2$ (square markers). Error bars indicate the 68$\%$ confidence intervals evaluated from a bootstrap analysis. Unless shown, error bars are smaller than the plotting symbols for the control sample. With the exception of low-redshift Radio AGN, all AGN hosts show an elevated level of star-formation activity, at all redshifts, with respect to their control sample of inactive galaxies.
\label{fig:sfa}}
\end{center}
\end{figure}
\vspace{0.5cm}

$UVJ$ diagnostics reveal that different AGN types (i.e. radio, X-ray or IR) are hosted by galaxies with different stellar properties at low redshift. This is consistent with studies that have examined the evolution of multi-wavelength AGN, where they're found to evolve with galaxies in the sequence of dusty IR AGN $\rightarrow$ unobscured X-ray AGN $\rightarrow$ early-type galaxy with intermittent radio AGN \citep{Hopkins:2006ju, Hickox:2009in, Goulding:2014ez}. This scenario is also supported by Figure~\ref{fig:sfa}, where our IR AGN exhibit a star-formation level consistent with young galaxies, radio AGN with quenched galaxies, and X-ray AGN straddling between the two. However, we find this trend weakens at higher redshifts ($z \gtrsim 0.8$), where all AGN are predominantly found to reside in star-forming galaxies, including our high-redshift radio AGN population. This being said, we remind the reader that a comparison between redshifts is inconclusive given the different luminosity depths used during AGN selection. Despite this, our result is supported by the recent findings of \citet{Rees:2015arXiv} who find the majority of radio AGN at $z > 1.5$ are hosted by star-forming galaxies. Such results contradict the before-mentioned perception that AGN hosts are traditionally viewed as quenched, elliptical galaxies.

As found in Figure~\ref{fig:uvj}, the $UVJ$ colours also reveal AGN hosts tend to be dustier than the control sample of galaxies. Similar to the offsets in star-formation, this is primarily driven by IR AGN, while for X-ray AGN the difference is marginal. These findings further support the scenario of an evolutionary sequence of dusty IR AGN $\rightarrow$ unobscured X-ray AGN, where copious amounts of gas and dust can fuel both a period of high star-formation and AGN before it begins to exhaust, star-formation slows, and X-rays from the AGN can shine through. Previous studies, which have examined star-formation activity in AGN hosts, commonly invoke a major merger scenario to interpret the finding of elevated star-formation over inactive galaxies \citep{Santini:2012jx, Rosario:2012hy, Juneau:2013kz}. In such a scenario, gas is driven to the central regions of merging galaxies, fuelling both a period of starburst and AGN activity. Merger driven elevations of star-formation activity has also been postulated to occur in Ultra-Luminous IR Galaxies and high-redshift submillimeter galaxies \citep{Pope:2013ig}.

Another possible explanation is that positive feedback from AGN activity triggers a flash of star-formation, which could lead to the elevated sSFRs seen in our AGN sample. A number of studies have shown observationally that AGN activity enhances star-formation in both radiatively efficient \citep[e.g.][]{Santini:2012jx} and inefficient AGN \citep[e.g.][]{Karouzos:2014fp} and is commonly explained by gravitationally collapsed cold gas resulting from AGN outflows, such as jets and accretion disk winds. Invoking this scenario would address the elevation of star-formation activity seen in our radio, X-ray and IR AGN hosts, but also leave the door open for the eventual quenching of star-formation from AGN negative feedback $-$ as seen in low-redshift radio AGN hosts.

While star-formation suppression is still required to reduce the over-predicted abundance of massive galaxies in models, we see no direct evidence AGN contributes to this suppression. Indeed, findings from recent simulations suggest that while AGN in isolated star-forming galaxies can remove substantial amounts of gas, this does not translate to a rapid quenching of star-formation \citep{Gabor:2014jx, Roos:2015bx}. Despite our inability to isolate a cause, the fact our AGN population exhibits a similar to slightly elevated level of star-formation activity over most of cosmic time $-$ and not a suppressed one $-$ calls into question the significance of AGN quenching as a major mechanism for moderating galaxy growth.

\section{Summary and Conclusion}
In this paper, we have utilised high-quality ground-based imaging from ZFOURGE in combination with ancillary data to select radio, X-ray and IR AGN hosts out to high redshifts ($z=0.2-3.2$). The deep imaging of ZFOURGE further provides us with host galaxy properties, including rest-frame colours, low stellar masses and accurate photometric redshifts. We maximise completeness by limiting our sample by mass, luminosity and redshift before conducting a detailed analysis of the rest-frame $UVJ$ colours and star-formation activity of AGN hosts. We also create a control sample of mass-matched, inactive galaxies in order to isolate the impact of AGN activity on star-formation. As discussed in Section~\ref{sec:con}, one of the uncertainties in this study (and all such studies) is conceivably the impact of AGN emission in the measurement of host galaxy properties. We assumed this impact is negligible, but it is difficult to test this assumption more rigorously. Our main findings are as follows:

\begin{enumerate}
  \item Radio, X-ray and IR-selected AGN hosts exhibit rest-frame $UVJ$ colours consistent with distinct galaxy populations. IR AGN tend to favour star-forming galaxies, radio AGN favour quiescent galaxies, and X-ray AGN straddle between the two. However, this distinction becomes blurred at higher redshifts ($z\gtrsim1.8$), where all AGN favour star-forming hosts. 
  \item The $UVJ$  diagnostics also reveal AGN have a higher dusty star-former fraction (${N}_{\mathrm{dusty}}/{N}_{\mathrm{sf}}$) and lower quiescent fraction (${N}_{\mathrm{q}}/({N}_{\mathrm{q}}+{N}_{\mathrm{sf}})$) when compared to the control sample of inactive galaxies.
  \item The star-formation activity (mean sSFR) of all AGN hosts tends to be elevated over inactive galaxies (average logarithmic offsets of $0.26\pm0.14$ dex at $z = 0.2-0.8$, $0.37\pm0.10$ dex at $z = 0.8-1.8$, and $0.38\pm0.10$ dex at $z = 1.8-3.2$).
  \item The star-formation activity (mean sSFR) of the split sample of radio, X-ray and IR AGN hosts is predominantly elevated over their respective control sample of inactive galaxies. IR AGN hosts exhibit an explicit and consistent $\sim 0.57$ dex elevation, X-ray AGN hosts a marginal $\sim 0.21$ dex elevation, while radio AGN hosts flip from a lower mean sSFR ($-0.53\pm0.20$ dex; $z = 0.2-0.8$) to higher level ($0.55\pm0.32$ dex; $z = 1.8-3.2$) at high redshift.
  \item One possibility for the elevated star-formation is that these AGN hosts are mergers where cold gas fuels both a period of starburst and AGN activity. Though not explored here, this scenario may be tested by comparing the morphologies determined from the existing HST imaging of these fields.
\end{enumerate}

Australian access to the Magellan Telescopes was supported through the National Collaborative Research Infrastructure Strategy of the Australian Federal Government. Additional scientific results are based in part on observations taken by the CANDELS Multi-Cycle Treasury Program with the NASA/ESA HST, operated by the Association of Universities for Research in Astronomy, Inc., under NASA contract NAS5-26555; XMM-Newton, an ESA science mission with instruments and contributions directly funded by ESA Member States and NASA;  the {\it Chandra} X-ray Observatory; and the National Radio Astronomy Observatory's Very Large Array, a facility of the National Science Foundation operated under cooperative agreement by Associated Universities, Inc. KG acknowledges support from ARC Grant DP1094370. SJ acknowledges support from the European Research Council via Stg-257720. GGK acknowledges from the Australian Research Council through the award of a Future Fellowship (FT140100933). We acknowledge support from Texas A\&M University and the George P. and Cynthia Woods Mitchell Institute for Fundamental Physics and Astronomy.




\bibliographystyle{mnras}
\bibliography{references} 


\bsp	

\begin{landscape}
\topskip0pt
\vspace*{\fill}
\begin{table}
\footnotesize
\centering 
\caption{ZFOURGE AGN Catalogue} 
\begin{tabular}{l c c c c c c c c c c c c c} 
\hline\hline
\noalign{\vskip 1mm}  
ID{\color{red}$^1$} & RA-J2000{\color{red}$^2$} & Dec-J2000{\color{red}$^3$} & $z_{\mathrm {phot}}${\color{red}$^4$} & $K{\mathrm{s}}_{\mathrm{mag}}$ {\color{red}$^5$} & $UVJ${\color{red}$^6$} & log($M_*$){\color{red}$^7$} & log($L_{\mathrm{UV}}$){\color{red}$^8$} & log($L_{\mathrm{IR}}$){\color{red}$^9$} & log($L_{\mathrm{x}}$){\color{red}$^{10}$} & log($L_{\mathrm{1.4GHz}}$){\color{red}$^{11}$} & IR-AGN{\color{red}$^{12}$} & X-AGN{\color{red}$^{13}$} & Rad-AGN{\color{red}$^{14}$} \\ [0.5ex]  
\hline  
\noalign{\vskip 1mm}  
1 & 150.16916 & 2.23389 & 1.233 & 21.6 & 2 & 10.00 & 10.18 & 11.44 & . . . & . . . & 1 & 0 & 0 \\[1ex]
2 & 53.10012 & -27.84268 & 1.361 & 22.6 & 2 & 9.58 & 10.02 & 11.49 & . . . & . . . & 1 & 0 & 0 \\[1ex]
3 & 34.39777 & -5.14525 & 1.479 & 21.6 & 3 & 10.77 & 10.15 & 11.82 & 43.77 & . . . & 0 & 1 & 0 \\[1ex]
4 & 150.04601 & 2.20114 & 0.922 & 21.8 & 2 & 10.71 & 11.43 & 12.57 & 44.42 & . . . & 0 & 1 & 0 \\[1ex]
5 & 53.08038 & -27.87200 & 1.102 & 21.1 & 1 & 10.81 & 9.45 & 10.41 & . . . & 23.19 & 0 & 0 & 1 \\[1ex]
6 & 34.35164 & -5.21450 & 0.906 & 20.1 & 1 & 11.14 & 9.78 & 11.02 & . . . & 25.10 & 0 & 0 & 1 \\[1ex]
7 & 150.13306 & 2.30325 & 1.712 & 21.3 & 3 & 11.44 & 10.50 & 12.53 & 44.96 & . . . & 1 & 1 & 0 \\[1ex]
8 & 150.06372 & 2.21119 & 1.486 & 21.5 & 3 & 10.75 & 9.58 & 12.34 & . . . & 24.43 & 1 & 0 & 1 \\[1ex]
9 & 53.05886 & -27.81949 & 2.050 & 23.3 & 3 & 9.80 & 10.34 & 11.67 & 42.14 & 24.13 & 0 & 1 & 1 \\[1ex]
10 & 150.16180 & 2.33236 & 1.256 & 21.4 & 3 & 10.52 & 9.88 & 11.39 & 43.73 & 23.68 & 1 & 1 & 1 \\[1ex]
\hline 
\noalign{\vskip 1mm}  
\multicolumn{14}{l}{%
\begin{minipage}{22.5cm}
Notes: This table is available in its entirety in a machine-readable form on the journal and ZFOURGE website: {\url{http://zfourge.tamu.edu}}. A portion is shown here for guidance regarding its form and content.
~\\
~\\  
Column 1: source ID number. Columns 2 and 3: J2000 RA and declination of the $K{\mathrm{s}}$-band selected hosts, respectively. Column 4: photometric redshift. Column 5: $K{\mathrm{s}}$-band magnitude (AB). Column 6: $UVJ$ criteria, where quiescent = 1, star-forming = 2 and dusty star-forming = 3. Column 7: host stellar mass (${M}_{\odot}$) Column 8: integrated 1216-3000\AA\ rest-frame UV luminosity ($L_{\odot}$). Column 9: integrated $8-1000$$\mu$m rest-frame IR luminosity ($L_{\odot}$). Column 10: 0.5-8 keV rest-frame luminosity ({erg s}$^{-1}$). Column 11: 1.4GHz rest-frame luminosity ($\mathrm{W\ Hz}^{-1}$). Columns 12-14: IR, X-ray and radio AGN flags, respectively, where AGN = 1, else = 0.
\end{minipage}
}\\
\end{tabular}
\label{table:catalog}
\end{table}
\vspace*{\fill}
\end{landscape}

\label{lastpage}
\end{document}